\documentclass[pra,twocolumn,superscriptaddress,showpacs]{revtex4}
\usepackage{graphicx}
\usepackage{amssymb}
\usepackage{epstopdf}
\usepackage{amsmath}
\usepackage{amstext}
\usepackage{latexsym}
\usepackage[matrix,frame,arrow]{xypic}

\newcommand{\abs}[1]{\left\vert #1 \right\vert}
\newcommand{\vet}[1]{\underline{#1}}
\newcommand{\one}{\mbox{$1 \hspace{-1.0mm}  {\bf l}$ }}  
\newcommand{\virgo}[1]{``#1''}
\newcommand{\pim}{\frac{\pi}{2}}
\newcommand{\fid}{\mathcal F}

\newcommand{\bra}[1]{\langle #1\vert}
\newcommand{\ket}[1]{\vert#1\rangle}

\newcommand{\arr}{\ar @{-} [r]}
\newcommand{\arx}[1][1]{\ar @{-} [#1,0]}

\newcommand{\wire}{\hspace{-10pt} \rule{6pt}{.4pt} \hspace{-10pt} \arr}
\newcommand{\gate}[1]{*+[F]{#1} \arr}

\newcommand{\control}{*+=[o]-[F]{\bullet}}

\newcommand{\targ}{*+=[o]=<3mm>[F]{+} \arr}
\newcommand{\multigate}[2]{{\phantom{#2}} \arr \save[0,0].[#1,0]!C *\frm{-} \restore}

\newcommand{\ghost}[1]{{\phantom{#1}} \arr}
\newcommand{\ctrl}[1]{\control \arx[#1] \arr}

\newcommand{\Qcircuit}{\xymatrix}

\begin{document}

\title{Cloning transformations in spin networks without external control}
\author{Gabriele De Chiara}
\affiliation{ NEST- INFM \& Scuola Normale Superiore, piazza dei
Cavalieri 7 , I-56126 Pisa, Italy}
\author{Rosario Fazio}
\affiliation{ NEST- INFM \& Scuola Normale Superiore, piazza dei
Cavalieri 7 , I-56126 Pisa, Italy}
\author{Chiara Macchiavello}
\affiliation{INFM \& Dipartimento di Fisica \virgo{A. Volta},
Via Bassi 6,I-27100 Pavia, Italy}
\author{Simone Montangero}
\affiliation{ NEST- INFM \& Scuola Normale Superiore, piazza dei
Cavalieri 7 , I-56126 Pisa, Italy}
\author{G. Massimo Palma}
\affiliation{ NEST- INFM \& Dipartimento di Tecnologie
dell'Informazione, Universita' degli studi di Milano\\ via
Bramante 65, I-26013 Crema(CR), Italy}
\date{\today}

\begin{abstract}
In this paper we present an approach to quantum cloning with
unmodulated spin networks. The cloner is realized by a proper
design of the network and a choice of the coupling between the
qubits. We show that in the case of phase covariant cloner the $XY$
coupling gives the best results. In the $1 \to 2$ cloning we find
that the value for the fidelity of the optimal cloner is achieved,
and values comparable to the optimal ones in the general $N \to M$
case can be attained. If a suitable set of network symmetries are
satisfied, the output fidelity of the clones does not depend on
the specific choice of the graph. We show that spin network
cloning is robust against the presence of static imperfections.
Moreover, in the presence of noise, it outperforms the
conventional approach. In this case the fidelity exceeds the
corresponding value obtained by quantum gates even for a very small
amount of noise. Furthermore we show how to use this method to
clone qutrits and qudits. By means of the Heisenberg coupling it
is also possible to implement the universal cloner although in
this case the fidelity is $10\%$ off that of the optimal cloner.

\end{abstract}

\maketitle

\section{Introduction}
\label{sec:intro} 
The no-cloning theorem~\cite{WZ82} states that
it is impossible to make perfect copies of an unknown quantum
state. At variance with the classical world, where it is possible
to duplicate information faithfully, the unitarity of time
evolution in quantum mechanics does not allow us to build a perfect
quantum copying machine. This no-go theorem is at the root of the
security of quantum cryptography~\cite{gisin02}, since an
eavesdropper is unable to copy the information transmitted through
a quantum channel without disturbing the communication itself.
Although perfect cloning is not allowed, it is, nevertheless,
possible to produce several approximate copies of a given state.
Several works, starting from the seminal paper by Bu\v{z}ek and M.
Hillery~\cite{BH96}, have been devoted to find the upper bounds to
the fidelity of approximate cloning transformations compatible
with the rules of quantum mechanics. Besides the theoretical
interest on its own, applications of quantum cloning can be found in quantum
cryptography, because they allow us to derive bounds for the security
in quantum communication~\cite{gisin02},  in quantum
computation, where quantum cloning can be used to improve the
performance of some computational tasks~\cite{galvao}, and in the problem 
of state estimation \cite{lopresti}.

As mentioned above, the efficiency of the cloning transformations
is usually quantified in terms of the fidelity of each output
cloned state with respect to the input. The largest possible
fidelity depends on several parameters and on the characteristics
of the input states. For an $N\to M$ cloner it depends on the
number $N$ of the input states and on the number $M$ of output
copies. It also depends on the dimension of the quantum systems to
be copied. Moreover, the fidelity increases if some prior
knowledge of the input states is assumed. In the universal cloning
machine the input state is unknown. A better fidelity is achieved,
for example, in the phase covariant cloner (PCC) where the state
is known to lie on the equator of the Bloch sphere (in the case of
qubits). Upper bounds to the fidelity for copying a quantum state
were obtained in Refs.~\cite{BH96} and ~\cite{bruss98} in the case
of universal and state dependent cloning respectively. The more
general problem of copying $N \to M$ qubits has been also
addressed~\cite{cloner_n-m}. The PCC has been proposed in Ref.
\cite{bruss00}. Several protocols for implementing cloning
machines have been already achieved
experimentally~\cite{cummins02,bouwmeester02,demartini,ekert03}.
In all the above proposals the cloning device is described in
terms of quantum gates, or otherwise is based on post-selection
methods. For example, the quantum network
corresponding to the $1\to 2$ PCC consists of two controlled-not (C-NOT) gates
together with a controlled rotation~\cite{niu99}.

The implementation of given tasks by means of quantum gates is not
the the only way to execute the required quantum protocols.
Recently it has been realized that there are situations where it
is sufficient to find a proper architecture for the qubit network
and an appropriate form for the coupling between qubits to achieve
the desired task. Under these conditions the execution of a
quantum protocol is reached by the time evolution of the
quantum-mechanical system. The only required control is on the
preparation of the initial state of the network and on the read-out 
after the evolution. This perspective
is certainly less flexible than the traditional approach with
quantum gates. Nevertheless it offers great advantages as it does
{\em not} require any time modulation for the qubits couplings.
Moreover, among the reasons for this \virgo{no-control} approach
to quantum information is that the system is better isolated from
the environment during its evolution. This is because there is no
active control on the Hamiltonian of the system. Actually, after
initializing the network one needs only to wait for some time (to
be determined by the particular task) and then read the output.
Several examples have been provided so far. A spin network for
quantum computation based only on Heisenberg interactions has been
proposed~\cite{benjamin03,yung}. Another area where this approach
is attracting increasing attention is quantum communication, where
spin chains have been proposed as natural candidates of quantum
channels~\cite{bose,subra03,datta,osborne,lloyd,cirac,giovannetti}.
An unknown quantum state can be prepared at one end of the chain
and then transferred to the other end by simply employing the
ability of the chain to propagate the state by means of its
dynamical evolution. These proposals seem to be particularly
suited for solid state quantum information, where schemes for
implementation have already been put
forward~\cite{romito,paternostro}.

Stimulated by the above results in quantum communication we have
studied quantum cloning in this framework. The main goal is to
find a spin network and an interaction Hamiltonian such that at
the end of its evolution the initial state of a spin is
(imperfectly) copied on the state of a suitable set of the
remaining spins. In this paper we will show that this is indeed
possible and we will analyze various types of quantum cloners
based on the procedure just described. We will describe a setup
for the $N\to M$ PCC and we will show that for $N=1$ and $M=2$ the
spin network cloning (SNC) achieves the optimal bound. We will
also describe the more general situation of cloning  of qudits,
i.e. d-level systems. An important test is to compare the
performance of our SNC with the traditional approach using quantum
gates. We show that in the (unavoidable) presence of noise our
method is far more robust. Some of the results have been
already given in Ref.~\cite{dechiara}.
In the present paper we will give many additional details, not contained 
in Ref.~\cite{dechiara},  and extend our approach to cloning to several 
other situations. We discuss cloning of qutrits, universal cloning machines, 
and optimization of the model Hamiltonian just to mention few extentions.

The paper is organized as follows. In Section \ref{sec:PCC} we
review the basic properties of approximate cloning showing the
theoretical optimal bounds. In Section~\ref{sec:SNC} we present
the models and the networks topologies considered in this work. In Sections 
\ref{sec:1Mpcc} and \ref{sec:NM} we briefly review and extend the results 
obtained in \cite{dechiara}. These sections concern the spin network model
to implement the $1\to M$ and $N\to M$ phase covariant cloning
transformations respectively.  In addition to a more detailed discussion, as 
compared to \cite{dechiara}, here we present a detailed analysis of the role 
of static imperfections. We also optimize the cloning protocol over the space 
of a large class of model Hamiltonians which includes the $XY$ and Heisenberg as 
limiting cases.

The effects of noise, included in a fully quantum mechanical approach, are 
analyzed in Section~\ref{sec:noise}, where we compare our cloning setup with
cloning machines based on a gate design. In Sect. \ref{sec:univ}
we study the possibility of achieving universal cloning with the
spin network approach. In Section \ref{sec:qudits} we generalize
the SNC for qutrits and qudits. Finally, in Section
\ref{sec:implementation} we propose a simple Josephson junctions
circuit that realizes the protocol and in \ref{sec:conclusion} we
summarize the main results and present our conclusions.

\section{Optimal fidelities for Quantum Cloning}
\label{sec:PCC}
Most of this  paper deals with the case of PCC.
We will therefore devote this section to  a brief summary
of the results known so far for the optimal fidelity achievable in this
case.

We start our discussion by considering quantum cloning of qubits,
whose Hilbert space is
spanned by the basis states $\ket{0}$ and $\ket{1}$.
The most general state of a qubit can be
parametrized by the angles $(\vartheta, \varphi)$ on the Bloch sphere as
follows
\begin{equation} \label{eq:psi}
\ket{\psi}=\cos\frac{\vartheta}{2}\ket{0}+
e^{i\varphi}\sin\frac{\vartheta}{2}\ket{1}\,.
\end{equation}
Quantum cloning was first analyzed \cite{BH96}, where the $1 \to
2$ universal quantum cloning machine (UQCM) was introduced. We
remind that the fidelity of a UQCM does not depend on
$(\vartheta, \varphi)$ i.e. it is the same for all possible input
states. As already mentioned, the quality of the cloner is
quantified by means of the fidelity $\mathcal F$ of each output
copy, described by the density operator $\rho$, with respect to
the original state $\ket{\psi}$
\begin{equation}
\mathcal F = \bra{\psi} \rho \ket{\psi}\;.
\end{equation}
The value of the optimal fidelity is achieved by maximizing $\fid$
over all possible cloning transformations. The result for the
$1\to 2$ UQCM is $\fid = 5/6\simeq 0.83$~\cite{BH96,bruss98}. The
general form of the optimal transformation, which requires an
auxiliary qubit, has been explicitly obtained in
Ref.~\cite{bruss98}.

When the initial state is known to be in a given subset of the
Bloch sphere, the value of the optimal fidelity generally
increases. For example, in Ref.~\cite{bruss98} cloning of just two non
orthogonal states is studied and it is shown that the fidelity in
this case is greater than that for the UQCM. The reason is that
now some prior knowledge information on the input state is
available. Another important class of transformations, which will
be largely analyzed in the present paper, is the so called phase
covariant cloning. In this type of cloner the fidelity is
optimised equally well for all states belonging to the equator of
the Bloch sphere:
 \begin{equation} \label{eq:psibloch}
\ket{\psi}=\frac{1}{\sqrt{2}}\left(\ket{0}+e^{i\varphi}\ket{1}\right )\;,
\end{equation}
where $\varphi\in[0,2\pi]$.
The optimal transformation for the $1 \to 2$ PCC was found in \cite{bruss00}.
The corresponding fidelity is given by
\begin{equation}
\fid = \frac{1}{2}+\frac{1}{\sqrt{8}} \simeq 0.854\;.
\end{equation}

In the $N \to M$ case of    PCC,
the optimal fidelities were derived in Ref. \cite{dariano03}.
For the $1 \to M$ PCC
case they read~\cite{footnote}
\begin{eqnarray}
\fid=&\frac{1}{2}\left ( 1+\frac{M+1}{2 M}\right ) & \quad \textrm{for odd} \, M\\
\fid=&\frac{1}{2}\left ( 1+\frac{\sqrt{M(M+2)}}{2 M}\right ) &\quad \textrm{for even} \, M
\end{eqnarray}

Optimal cloning has also been studied in higher dimensions. The
universal case was discussed in Refs.~\cite{werner98}. Cloning of
qutrits was specifically treated in
Refs.~\cite{lopresti,gisinqutrit}, where the optimal fidelity for
cloning some classes of states was derived. In particular the
double-phase covariant symmetric $1\to 2$ cloner , which is
optimized for input states of the form
\begin{equation} \label{doublePCC}
\ket \psi =\frac{1}{\sqrt 3} \left(\ket 0 +e^{i \varphi_1}\ket 1+e^{i \varphi_2}\ket 2 \right)
\end{equation}
where $\varphi_1$ and $\varphi_2$ are independent phases and the
states $\ket 0, \ket 1 , \ket 2$ form a basis for a qutrit, was
analyzed \cite{lopresti}. The optimal $1\to M$ fidelity in the
case of $M=3k+1$ (with $k$ positive integer) is given by the
simple expression \cite{dariano03}:
\begin{equation}
\fid=\frac{1}{3}\left( 1+ 2 \frac{M+2}{3M} \right)
\end{equation}

For qudits, namely quantum systems with finite dimension $d$, the
PCC is optimized for states of the form
\begin{equation}
\ket \psi =\frac{1}{\sqrt d}\sum_{i=0}^{d-1} e^{i\varphi_i} \ket i
\end{equation}
where $\varphi_0=0$, while $\varphi_1,\varphi_2,\dots,\varphi_{d-1}$ are
independent phases.
The fidelity for $1\to 2$ PCC in dimension $d$ is given by \cite{matsu}
\begin{equation}
\fid = \frac{1}{d} +\frac{1}{4 d}\left(d-2+\sqrt{d^2+4 d-4}\right)\;.
\end{equation}

More recently the general $N\to M$ case of PCC for qudits was analysed
\cite{bdm}, where explicit simple solutions were obtained for a number of output
copies given by $M=kd+N$, with $k$ positive integer.

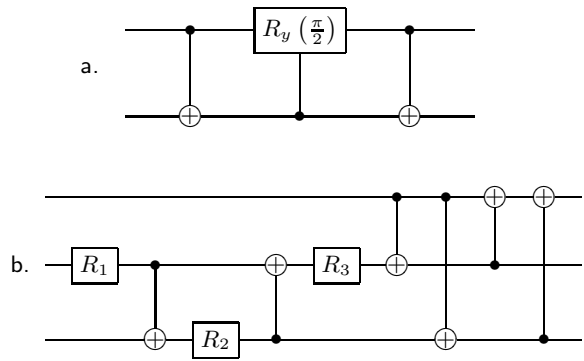
\begin{figure}[htbp]
{\sf a.}
\parbox{5 cm}{
\begin{displaymath}
  \Qcircuit @C=20pt @R=20pt {
    \arr &\ctrl{1}&\gate{R_y\left(\pim\right)}&\ctrl{1}&\\
    \arr & \targ  & \ctrl{-1}                          &\targ   &
                            }
\end{displaymath} }
\\
 {\sf b.}\parbox{7 cm}{\begin{displaymath}
  \Qcircuit @C=10pt @R=15pt {
   \arr &\wire            &\wire            & \wire   &\wire &\wire           &\ctrl{1}&\ctrl{2}&\targ    &\targ&\\
    \arr & \gate{R_1}&\ctrl{1}          & \wire   &\targ &\gate{R_3} & \targ  & \wire  &\ctrl{-1}&\wire&\\
    \arr & \wire&\targ  & \gate{R_2}&\ctrl{-1}& \wire&\wire           &\targ   &\wire   &\ctrl{-2}&
                            }
\end{displaymath}
   }
  \caption{{\sf a.} Circuit implementing $1\to 2$ PCC for qubits and experimentally realized in \cite{du}. {\sf b.} Circuit implementing $1 \to 3$ PCC \cite{buzek97}. }
  \label{fig:circuitPCC}
\end{figure}

The quantum transformations corresponding to the cloning machines
described above can be implemented by suitably designed quantum
circuits. For example, in the $1 \to 2$ and $1 \to 3$ cases the
circuits implementing PCC were derived in Refs. \cite{du} and
\cite{buzek97} respectively and they are shown in
Fig.\ref{fig:circuitPCC} (note that in these cases no auxiliary
qubits are needed).
In Fig.\ref{fig:circuitPCC}{\sf b.} we defined $R_i=R_y(-2\vartheta_i)$ and
$\vartheta_1=\vartheta_3=\pi/8$ and $\vartheta_2=\arcsin\sqrt{\frac{1}{2}-\frac{\sqrt{2}}{3}}$.

This is not the only way to perform quantum cloning and quantum computation
protocols in general.
In the following sections we describe how to accomplish the desired task 
using the evolution of a spin network. The expressions for the fidelities 
given above will be compared to the ones derived by means of the approach 
proposed in Ref.~\cite{dechiara} and in the present paper.

\section{The spin network cloning}
\label{sec:SNC}

In this section we show how quantum cloning can be implemented
using a spin network. First let us discuss our Hamiltonian model.
We start with a fairly general model defined as
\begin{equation}
        H_{\lambda} = \frac{1}{4} \sum_{ij} J_{ij}(\sigma^i_x \sigma^j_x +
        \sigma^i_y \sigma^j_y
        + \lambda \sigma^i_z \sigma^j_z ) + \frac{B}{2} \sum_{i}  \sigma^i_z
\label{eq:hamiltonian}
\end{equation}
where $\sigma^i_{x,y,z}$ are the Pauli matrices corresponding to
the $i$-th site, $J_{ij}$ are the exchange couplings defined on
the links joining the sites $i$ and $j$ and $B$ is an externally
applied magnetic field. The anisotropy parameter $\lambda$ ranges
from $0$ ($XY$ Model) to $1$ (Heisenberg Model). This model is named
as the $XXZ$ Model. We discuss separately the two limiting cases
$\lambda =1$  and  $\lambda=0$. It turns out that for PCC the
$\lambda =0$ case leads always to the highest fidelity. Given the
model of Eq. \eqref{eq:hamiltonian} the fidelity is maximized over
$B/J$ and $J t$. We defined $B^{(M)}$ and $t^{(M)}$ the values of
the parameters leading to the optimal solution. Notice that the
total angular momentum as well as its $z$ component are always
constants of motion independently of the topology of the network.
In all the cases we consider in this work, the couplings $J_{ij}
\neq 0$ only for nearest neighbors sites $i,j$. To specify which
couplings are non-zero one has to define the topology of the spin
network.

\begin{figure}[htbp]
  \centering
 \includegraphics*[scale=0.4]{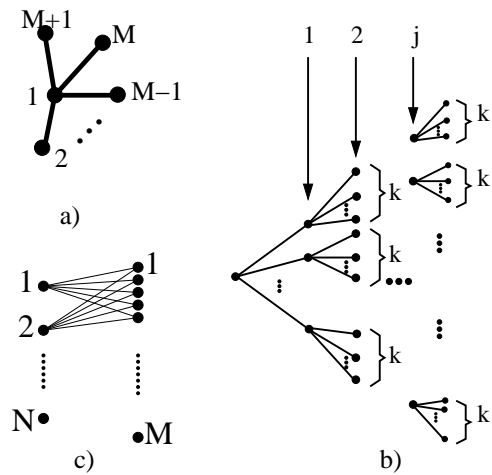}
        \caption{Different topologies for $N \to M$ cloner:
 a) Spin star network for $1 \to M$ cloner. b) Generic graph for the $1
        \to M$ cloner with $j$ intermediate steps and $k$ links
        departing from each vertex. c) Spin network for the $N \to M$ cloning}
      \label{fig:network}
\end{figure}

For the $1\to M$ PCC we choose $M+1$ spins in a star configuration
(see fig. \ref{fig:network}{\sf a}). The central spin labeled by
$1$ is initialized in the input state while the remaining M spins
are the blank qubits and are initialized to the state $\ket{0}$ if
$0<\vartheta<\pi/2$ and $\ket{1}$ if $\pi/2<\vartheta<\pi$. For
this network $J_{ij}=J$ only if one of the two linked sites is the
central one. This configuration has also been studied in a
different context~\cite{Bosestar}. For the $1\to M$ case we have
considered also other types of networks. These are represented in
Fig.\ref{fig:network}{\sf b}. The original state is placed on the
top of the tree while the blank qubits are on the lowest level.
The intermediary spins are ancillae. Each graph is characterized
by the number $k$ of links departing from each site and the number
$j$ of intermediate levels between the top and the blank qubits
level. The number of blank qubits can be obtained from $j$ and $k$
as $M=k^{j+1}$. Notice that for this class of graphs a symmetry
property holds: the global state of the blank qubits is invariant
under any permutation of the $M$ qubits. For the $N\to M$ PCC we
have considered a generalization of the star network (see
Fig.\ref{fig:network}{\sf c}). It consists of a star with N
centers and M tips so no auxiliary qubits are present. Also this
network is permutation invariant.

The cases defined above are not the most general networks and/or
model Hamiltonian conceivable.  Since the fidelity must be
maximized over the parameters of the Hamiltonian as well as over
the network topology one may wonder whether it is sufficient to
consider only the cases introduced above. In Section
\ref{sec:maxham} we partially answer to this question by
considering the more general configuration for the $1\to 3$ case
containing $4$ spins and 
we believe to have found the best possible scenario for the SNC.
As far as the choice of the model Hamiltonian is concerned, the
symmetry in the $XY$-plane is suggested by the phase covariance
requirement for a PCC. We checked that the Ising model in
transverse field, which is not phase covariant, gives poorer results
for the fidelity. In principle one should also explore the
possibility of multi-bit couplings, but we did not considered this
(in principle interesting) situation. Multi-bit couplings are much
more difficult to achieve experimentally and at the end of this
work we want to propose to implement our scheme using Josephson
nanocircuits, where two-qubit couplings with $XY$-symmetry are easy
to realize.

\section{$1\to M$ PCC cloning}
\label{sec:1Mpcc} In this case the Hamiltonian
\eqref{eq:hamiltonian} can be easily diagonalized because it can
be mapped to the problem of a spin-1/2 interacting with a spin-M/2
\begin{equation}
H_{\lambda}=J\left(S^1_xS_x+S^1_yS_y+\lambda S^1_zS_z\right) +B(S^1_z +S_z) \;\; .
\end{equation}
where we have defined $S^1=\sigma^1/2$ and $S=1/2\sum_2^{M+1}
\sigma^i$. The operators $S$ obey the usual commutation relations
for an angular momentum operator. Notice that the modulus and the
$z$ component of the total angular momentum $\vec F = \vec S^1+ \vec
S$  commute with the Hamiltonian. Thus the evolution is invariant
under rotations around the $z$ axis. This property automatically
makes our model a PCC.

For $0<\vartheta<\pi/2$ (the other case is equivalent) the initial state of the star is:
\begin{equation}
\ket{\Psi(0)}=\alpha\ket{\vet 0}+\beta\ket{\vet 1}
\end{equation}
where $\alpha=\cos\frac{\vartheta}{2}$ and $\beta=e^{i\varphi}\sin\frac{\vartheta}{2}$.
Here we use the convention that $\ket{\vet 0}=\ket{\underbrace{0 \cdots 0}_{M+1}}$
and $\ket{\vet j}=\ket{\underbrace{0\cdots 1\cdots 0}_{M+1}} $ where only the site $j$th
is in the state $\ket{1}$.

Because of the conservation of the total angular momentum the state of the star at
time $t$ will be a linear combination of $\ket{\vet 0}$ and $\ket{\vet j}$ i.e.  states
with only one qubit in state $\ket{1}$.
The state at time $t$ can thus be written in the form (apart
from a global phase factor):
 \begin{equation}  \label{eq:Psit}
   \ket{\Psi(t)} =
                \alpha \ket{\vet{0}} + \beta_1(t)\ket{\vet{1}}
                   +\beta_2(t)\frac{1}{\sqrt{M}}\sum_{j=2}^{M+1}
        \ket{\vet{j}}
\end{equation}
where the coefficients $\beta_1(t)$ and $\beta_2(t)$ depend on the
particular choice of the Hamiltonian. In order to calculate the
fidelity of the clones we need the expression for the reduced
density matrix of one site. For symmetry reasons this is
independent on the site chosen, the result being
\begin{equation} \label{eq:rho}
  \rho(t) = \left(\begin{array}{cc}
        |\alpha|^2 + |\beta_1|^2 +\left(1-\frac{1}{M}\right) |\beta_2|^2 &
                \frac{\alpha\beta_2^*}{\sqrt{M}}   \\
                \frac{\alpha^*\beta_2}{\sqrt{M}}  &
                \frac{|\beta_2|^2}{M}
              \end{array}
                  \right) \;\; .
\end{equation}
The fidelity for the SNC is
\begin{eqnarray}
\mathcal{F}_{\lambda} &=& |\alpha|^2\left[|\alpha|^2 + |\beta_1|^2 +
                              \left(1-\frac{1}{M}\right)|\beta_2|^2 \right]
        \nonumber \\
        &+&
             \frac{|\beta_1 \beta_2|^2}{M}
        +
2\textrm{Re} \left[ \frac{|\alpha|^2 \beta_1^* \beta_2}{\sqrt{M}}\right]
\end{eqnarray}
where the coefficients $\beta_i(t)$ depend explicitly on the
chosen model (Heisenberg or $XY$ in this case).

\underline{Heisenberg model -}
Let us start with the Heisenberg model ($\lambda=1$) and let $B=0$ (we checked that a
finite external magnetic field is not necessary to achieve the maximum fidelity).
The Hamiltonian can be rewritten in the form $H=J\vec S^1 \cdot\vec S$ and using
$\vec{S^1} \cdot \vec{S} = 1/2 (F^2 - S^2 - S_1^2)$ one finds that the eigenenergies are
given by $E(F,S,S^1) = \frac{J}{2} \left[ F(F+1) - S(S+1)-S^1(S^1+1) \right]$ where
$F,S,S^1$ are the quantum numbers associated to the corresponding operators.
The eigenvectors can be found in terms of the Clebsch-Gordan coefficients.
The results for the coefficients $\beta_i(t)$ are:
 \begin{eqnarray}  \label{hbetat}
\beta_1(t) &=&\beta \frac{1}{1+2S} \left[2S\, e^{i\left(\frac{1}{2}+S\right)t} +1\right]
                                                     \label{hbeta1} \\
\beta_2(t) &=& \beta\frac{\sqrt{2S} }{1+2S}\left[1- e^{i\left(\frac{1}{2}+S\right)t}\right]
                                                     \label{hbeta2}
 \end{eqnarray}
where $S=M/2$.

The maximum value for the fidelity
\begin{equation}
\mathcal{F}_{1}=\frac{4 +
     M\,(3 + M)  +
     (M-1)[( 3 + M )\cos \vartheta -
     \cos 2\vartheta ]}{2
     {( 1 + M ) }^2}
\end{equation}
is obtained for the parameters $J t^{(M)}=2\pi /(M+1)$.

\begin{figure}[t!]
\begin{center}
\includegraphics[scale=0.3]{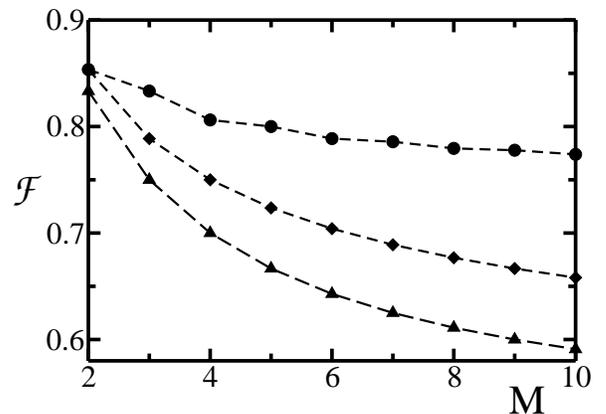}
\end{center}
  \caption{The fidelity $\mathcal F$ for PCC (circle), $XY$
          (diamond) and Heisenberg (triangle) as functions of $M$ for
          $\vartheta=\pi/2$.}
  \label{fig:1mcloning}
\end{figure}

\underline{$XY$ model -} Now let us turn our attention to the $XY$ model.
Solving the eigenvalue problem as in \cite{Bosestar} one finds
\begin{eqnarray}
\beta_1(t) &=& \beta e^{iBt} \cos \frac{J}{2}\sqrt{M} t \\
 \beta_2 (t) &=& -i\beta e^{iBt} \sin \frac{J}{2}\sqrt{M} t \\
\end{eqnarray}
The fidelity is maximized when $B^{(M)}/J=\sqrt{M}/2$ and $J t^{(M)} =\pi/\sqrt{M}$:
\begin{equation}
\mathcal F_{0}=\frac{(1 + {\sqrt{M}})^2 -
    \left(2 - 2 M \right)
     \cos \vartheta  +
    \left(1 - {\sqrt{M}} \right)
     \cos 2\vartheta }{4M}
\end{equation}
\begin{figure}[t!]
\begin{center}
\includegraphics[scale=0.3]{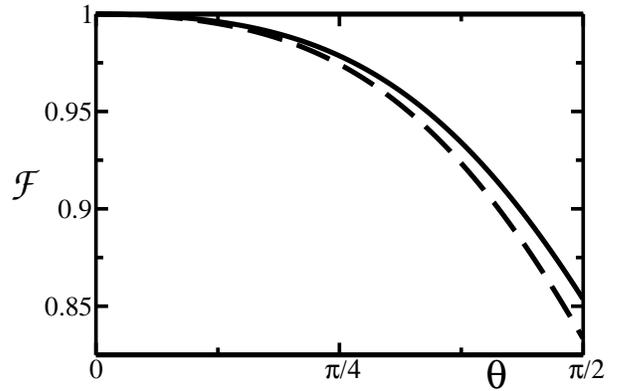}
\end{center}
    \caption{The fidelity $\mathcal F_\lambda$ for $M=2$ versus $\vartheta$ for
        the $XY$ (solid) and Heisenberg (dashed) model are
        shown. Notice that the optimal fidelity for the PCC is exactly
        that of the $XY$ model.}
  \label{fig:12cloning}
\end{figure}
For $(\lambda=0)$ the fidelity is always greater than for
$(\lambda=1)$. Let us analyze the previous results in two important
cases. First let us discuss the case  $M=2$ for arbitrary
$\vartheta$ (see Fig.\ref{fig:12cloning}):
\begin{eqnarray}
\mathcal{F}_{1} &=& \frac{14 + 5\cos\vartheta - \cos 2\vartheta}{18} \\
\mathcal{F}_{0} &=& \frac{1}{8}(5 + 2\cos \vartheta + \cos 2\vartheta  +
    2{\sqrt{2}}\sin^2 \vartheta)
\end{eqnarray}
The fidelity $\mathcal F_{0}$ {\em coincides} with the fidelity
for the $1 \to 2$ PCC~\cite{fiurasek03} i.e. the SNC saturates the
optimal bound for the $1\to 2$ PCC.
Second let us consider
$\vartheta=\pi/2$ and arbitrary M (see Fig.\ref{fig:1mcloning}):
\begin{eqnarray} 
\fid_{1}  &=& \frac{1}{2} + \frac{1}{1+M} \label{eq:fh1m}\\
\fid_{0}    &=& \frac{1}{2}\left(1 + \frac{1}{\sqrt{M}} \right) \label{eq:fxy1m}
\end{eqnarray}
For $M>2$,  $\mathcal F_\lambda$ is always smaller than the
optimal fidelity given in Ref.\cite{dariano03}. Also in this case
the XY model is better suited for quantum cloning as compared to
the Heisenberg case. Although for generic $M$ the SNC does not
saturate the optimal bound, there is a very appealing feature of
this methods which makes it interesting also in this case. The
time required to clone the state {\em decreases} with $M$. This
implies that, in the presence of noise, SNC may be competitive
with the quantum circuit approach, where the number of gates are
expected to {\em increase} with $M$. We analyze this point in
Section \ref{sec:noise}.

Recently a PCC with the star configuration has been proposed also 
for a multi-qubit cavity \cite{olayacastro}. In this proposal the 
central spin is replaced by a bosonic mode of the cavity. By 
restricting the dynamics in the subspace with only one excitation 
(one excited qubit or one photon in the cavity) the Hamiltonian is 
equivalent to the $XY$ spin star network considered here. 
Indeed  the optimal fidelities coincide with $\fid_0$, 
Eq.\eqref{eq:fxy1m}.

All the results discussed so far have been obtained for the star
network. Obviously this is not the only choice which fulfills the
symmetries of a quantum cloning network. In general one should
also consider more general topologies and understand to what
extent the fidelity depends on the topology. We analyzed this
issue by studying the fidelity for the $XY$ model and $\vartheta=
\pi/2$  for $M \le 32$ for the graph b of Fig.\ref{fig:network}
(the fidelity for Heisenberg model in this case is much worse than
in the star configuration). We conclude that the maximum fidelity
obtained does not depend on the chosen graph.

\subsection{Imperfections}
\begin{figure}[htbp]
  \centering
  \includegraphics*[scale=0.3]{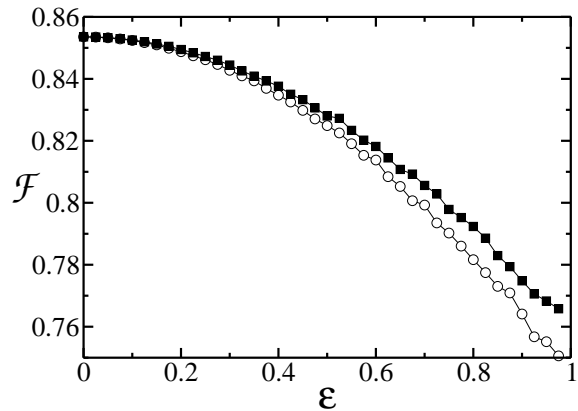}
        \caption{Mean fidelity for the $1\to 2$ case with static imperfections
 as a function of the tolerance $\epsilon$ with $\mu=0$ (filled squares) and $\mu=0.5$ (empty circles). }
      \label{fig:imper}
\end{figure}

To assess the robustness of our protocol, it is important to
analyze the effect of static imperfections in the network. In a
nanofabricated network, as for example with Josephson
nanocircuits, one may expect small variations in the qubit
couplings.  Here we analyze the $1 \to M$
cloning assuming that the couplings $J_{ij}$ have a certain degree
of randomness. For each configuration of disorder $J_{1i}$ are
assigned in an interval of amplitude $2\epsilon$ centered around
$J=1$ with a uniform distribution. First we study the case of uncorrelated disorder in different links. The values of $B$ and $t$ are
chosen to be the optimal values of the ideal situation. For a
given configuration of the couplings the fidelities of each of the
clones are different due to the different coupling with the
central spin. Only the average fidelity is again symmetric under
permutation among the clones. We averaged the fidelity over the
$M$ sites and over $10^3$ realization of disorder. For
$\epsilon=10^{-1}$ and $M \le 10$ the mean fidelity decreases by
just less than $0.2\%$  of the optimal value. It is important to
stress that the effect of imperfections is quite weak on the
average fidelity. This is because for certain values of $J_{ij}$,
even if the fidelity of a particular site can become much larger
than the fidelity in the absence of disorder, at the same time for
the same parameters the fidelity in other sites is very small and
the average fidelity is weakly affected by imperfections. In
figure \ref{fig:imper} we show the fidelity for the $1\to 2$ SNC
with imperfections as a function of the tolerance $\epsilon$.
We study also the case with correlations between the signs of nearest neighbor bonds: the probability of equal signs $(J_{1i}-J)(J_{1i+1}-J)>0 $ is proportional to $\mu \in [-1;1]$.
The uncorrelated results are recovered for $\mu=0$. As expected this type of disorder is more destructive as shown in figure \ref{fig:imper}.

\subsection{Optimal network Hamiltonian for $1 \to 3$ PCC}
\label{sec:maxham} 
\begin{figure}[htbp]
\begin{center}
\includegraphics[scale=0.5]{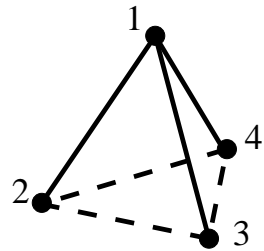}
\end{center}
        \caption{The tetrahedron network analyzed for the $1 \to 3$ cloning.}
      \label{fig:tetra}
\end{figure}

As shown in Fig.\ref{fig:1mcloning}, the $1\to 2$ SNC saturates the
PCC optimal bound. However this is not the case for $M > 2$, at
least for the network topologies considered up to now. One may
wonder whether a different choice for the network could allow to
approach the optimal fidelity.  In order to understand this point
we studied the simplest non trivial case namely $M=3$ and
considered the tetrahedron network shown in Fig.\ref{fig:tetra}.

We concentrated on the general anisotropic $XXZ$ model presented in
\eqref{eq:hamiltonian} in which the local magnetic field and the
couplings between the central spin and the blank spin can be
different. This is the most general Hamiltonian for $4$ spins that
fulfills the symmetry and covariance property. For this general
model we maximized analytically the on-site fidelity diagonalizing
the corresponding Hamiltonian. We found that the maximum fidelity
exactly coincides with that found with the simple star
configuration. It is thus demonstrated that, at least for $M \le
3$, the star configuration is the optimal network for cloning.

It is however important to stress that, given the transformation
for the optimal PCC, it is always possible to find a Hamiltonian
that generates this transformation during the dynamical evolution.
Therefore, at least in principle, one should be able to saturate
the optimal value by including other terms in the Hamiltonian
(multi-spin coupling for example). On purpose we chose to limit
ourselves to a fairly general model which however can be realized
experimentally.

\section{$N\to M$ PCC cloning}
\label{sec:NM}

In this section we discuss the generalization of the SNC to the
$N>1$ case. The suitable network  to accomplish this task is
depicted in Fig.\ref{fig:network}{\sf c}. The model can be mapped
to the problem of the interaction between two higher dimensional
spins, $N/2$ and $M/2$ respectively. Since we did not succeed in
finding the analytic solution to the problem (for example for
$2\to 8$ the relevant subspace has dimension $56$), we simulated
it numerically.
\begin{table}
  \begin{tabular}{|c|c|c|c|c|c|c|}
\hline
 $N$ &$M$ &  $\mathcal F_{PCC}$ &$\mathcal F_{abs}$&$J t_c$ &$J/B$ &$J t_c(10^{-2})$\\
\hline
2&3& 0.941& 0.938&1516.0 & 39.5&2.9\\
\hline
2&4& 0.933& 0.889&53.1 &4.5&5.3\\
\hline
2&5& 0.912&0.853  & 774.1& 12.8&2.7\\
\hline
2&6& 0.908&0.825&563.4 &28.7&2.8\\
\hline
2&7& 0.898 & 0.804 &156.0 &40&4.9 \\
\hline
2&8& 0.895 & 0.786 &116.6 &29.9&6.9\\
\hline
3&4& 0.973 &0.967 &2201.6 &47.5&111.8\\
\hline
3&5& 0.970 & 0.931&1585.5 &33.1&19.6\\
\hline
3&6& 0.956&0.905 &8.3 &10.6&8.3\\
\hline
3&7&0.954 &0.875 &8.1 &3.4&7.9\\
\hline
  \end{tabular} 
  \caption{The maximum fidelity $\mathcal F$ for $N \to M$ for the
 network of Fig.\ref{fig:network}c. $\mathcal F_{PCC}$ is the optimal
 fidelity for the PCC \cite{dariano03}. Column 5 (6) reports the
 corresponding evolution time $t_c$
  (interaction strength $J$). Column 7 reports the time $t_c(\epsilon=10^{-2})$
 at which the fidelity reaches the value $\fid_{abs}-10^{-2}$. The results 
refer to the $XY$ model
 ($\lambda=0$).
The value $\mathcal F$ is found by numerical maximization in the
 intervals $B/J \in [0.01;10]$ for $N+M<10$  and $J t \in [0;5 \cdot 10^3]$.}
  \label{tab:nmcloning}
  \end{table}
We have simulated the evolution of the network in the range $B/J \in [0.01;10]$ 
and $tJ<
5\cdot 10^3 $. We found the absolute  maximum of the fidelity $\fid_{abs}$ in
this interval. The result of this maximization is summarized in
Table \ref{tab:nmcloning} for several values of $N$ and $M$. We also calculated
 the time to reach a value of fidelity slightly lower than $\fid_{abs}$. The 
time needed to reach $\fid_{abs}-\delta$, $\delta \ll 1$,  is greatly 
reduced. Indeed the fidelity is a quasi periodic function of time approaching 
several times values very close to $\fid_{abs}$. In Table
\ref{tab:nmcloning} both the absolute maximum $\fid_{abs}$ (column
4) in the chosen interval and the time $t_c(\delta=10^{-2})$ (last column in 
the table) are shown.


\section{Quantum cloning in the presence of noise}
\label{sec:noise}

\begin{figure}[htbp]
  \begin{center}
     \includegraphics[scale=0.3]{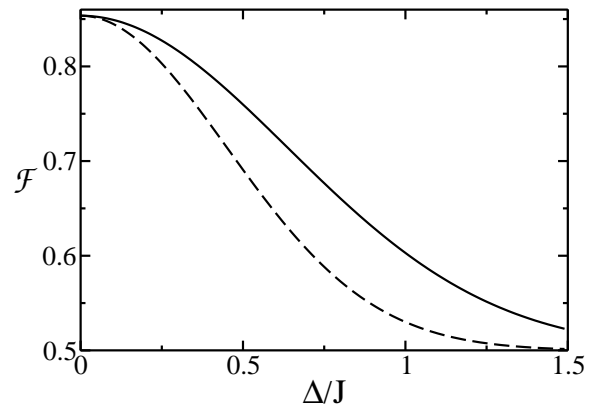}
  \caption{Fidelity for equatorial qubits in the $XY$ model with a  classical
    fluctuating field. $\fid$ is plotted as a function of the variance
    $\Delta$ for fluctuating $J$ (solid) and $B$ (dashed).}
  \label{clnoise}
  \end{center}
 \end{figure}

So far we have described the unitary evolution of isolated spin
networks. Real systems however are always coupled to an
environment which destroys their coherence. In this section we
will try to understand the effect of noise on the SNC. We will
also compare the performances of quantum cloning machines
implemented with spin networks and with quantum circuits using the
same Hamiltonian. The effect of the environment can be modeled in different
ways. One is to add classical fluctuations to the
external magnetic field $B$ or the coupling $J$. These random fluctuations 
can be either time  independent or
stationary stochastic processes. In both cases one can define an
effective field variance $\Delta$ and average the resulting
fidelity. In Fig. \ref{clnoise} we compare the fidelity
$\fid_{1\to 2}$ and $\vartheta=\pi/2$ as a function of $\Delta$
for the $XY$-model with the optimal average values for fluctuating
$J$ (solid) and $B$ (dashed). The probability distributions are
chosen to be Gaussian. Note that the fidelity is more sensitive to
fluctuations of $B$. \vskip 1 cm

However there are situations in which the environment cannot be
modeled as classical noise and one has to use a fully quantum
mechanical description. Following the standard approach, we model
the effects of a quantum environment by coupling the spin network
to a bosonic bath. Then we describe the time evolution for the
reduced density matrix of the spin system alone, after tracing out
the bath degrees of freedom in terms of a master
equation~\cite{cohen}. The Hamiltonian for the whole system is
\begin{eqnarray}
  \label{mehtot}
  H &=& H_S + H_{R} +H_I \\
  H_I &=&\sum_{i=1}^{M+1}\sum_{k} \lambda_i(k)\,\, \sigma_z^i
  \left[a_i^\dagger(k)+a_i(k)\right] \\
  H_{R}&=&\sum_{i=1}^{M+1}\sum_{k} \omega_i(k)\,\, a_i^\dagger(k)\,a_i(k)
\end{eqnarray}
where $H_S$ is the spin Hamiltonian defined in Eq.
\eqref{eq:hamiltonian}. The model is presented for generic $M$ but
we will discuss the results only for $M=2$ and $M=3$. We suppose
that each spin is coupled to a different bath, labeled by $i$, and
that all baths are independent, $\omega_i(k)$ and $\lambda_i(k)$
are the frequency and the coupling constant of the $k$th mode of
the $i$th bath. It is convenient to define the operator
$E_i=\sum_k \lambda_i(k) \left[a_i^\dagger(k)+a_i(k)\right] $, the
environment operator to which the system is coupled.

The master equation in the basis of eigenstates of $H_S$ can be written as:
\begin{equation}
\frac{d}{dt}\rho_{ab}= - \sum_{abcd}  \mathcal R_{abcd} \, \, \rho_{cd}
\end{equation}
where the indexes $a,b,c,d$ run over the energy eigenstates and
$\mathcal R_{abcd}$ is the so called Bloch-Redfield tensor in the interaction
picture:
\begin{equation}
{\cal R}_{abcd} = \sum_i \int_{0}^{\infty}d\tau
    \left\{ G_i(\tau ) \Sigma_{abcd}^{>}
    + G_i(-\tau )\Sigma_{abcd}^{<} \right\}
\end{equation}
where
\begin{equation}
\Sigma_{abcd}^{>}=
    \delta_{bd}\sum_n (\sigma_z^i)_{an}(\sigma_z^i)_{nc}e^{i\omega_{cn}\tau}-
    (\sigma_z^i)_{ac}(\sigma_z^i)_{db}e^{i\omega_{ac}\tau}
\end{equation}
and
\begin{equation}
    \Sigma_{abcd}^{<}=
    \delta_{ac}\sum_n (\sigma_z^i)_{dn}(\sigma_z^i)_{nb}e^{i\omega_{nd}\tau} -
    (\sigma_z^i)_{ac}(\sigma_z^i)_{db}e^{i\omega_{bd}\tau}
\end{equation}
with $(\sigma_z^i)_{ab} = \langle a \vert \sigma_z^i \vert b\rangle$.
The function $G(\tau )$ is the correlation function of the environment
operators in the interaction picture:
\begin{equation}
G_i(\tau ) = \text{Tr}\left[ \rho_F\widetilde{E_i}(\tau
)\widetilde{E_i}(0) \right]
\end{equation}
The functions $G_i(\tau)$ can be related to the spectral density of the bath through
\begin{equation} \label{eq:gomega}
\left[ G_i(\tau) \right]_\omega=2 N_i(\omega) J_i(\omega)
\end{equation}
where $\left[\cdot \right ]_\omega$ indicates the Fourier transform. In
Eq.\eqref{eq:gomega} $N_i(\omega)=(e^{\beta\omega}-1)^{-1}$ is the mean
occupation number of the $\omega_i$ mode at temperature $T=\beta^{-1}$ and
$J_i(\omega)=\pi\sum_{\omega_k}  \abs{\lambda_i(\omega_k)}^2 \delta(\omega-\omega_k)$
is the spectral density. We suppose that the bath is Ohmic, as often encountered in
several situations, i.e. $J(\omega)$ has a simple
linear dependency at low frequencies up to some cut-off:
\begin{equation}
  \label{spectraldensity}
  J_i(\omega)=\frac{\pi}{2} \alpha \omega e^{-\omega/\omega_C}
\end{equation}
The parameter $\alpha$ represents the strength of the noise and $\omega_C$
is the cut-off frequency.

\begin{figure}[htbp]
\vspace{0.5 cm}
\begin{center}
\includegraphics[scale=0.3]{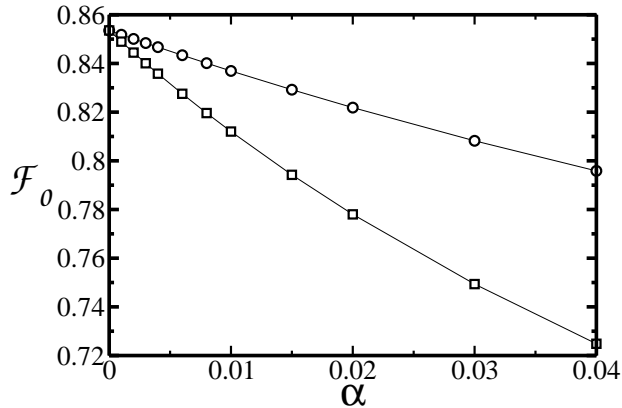}
\end{center}
  \caption{$1 \to 2$ cloning.  Comparison of the fidelity $\mathcal F_0$ obtained by the spin
    network method and the quantum circuit ($XY$ interaction) discussed
    in \protect\cite{du,buzek97} in the presence of an external quantum noise.
    Circles and squares refer to  the network and gates case respectively
    ($\vartheta=\pi/2$.).
    The parameters for the environment are $\beta=10/J$ and $\omega_C=10^4 J$.}
  \label{fig:beta10a}
\end{figure}

\begin{figure}[htbp]
\vspace{0.5 cm}
\begin{center}
\includegraphics[scale=0.3]{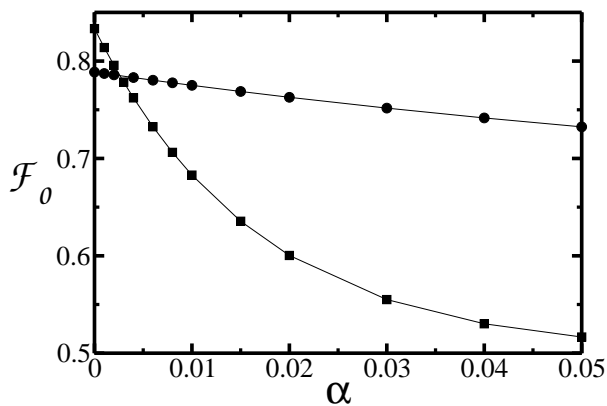}
\end{center}
  \caption{The same as in Fig.\ref{fig:beta10a} for the $1 \to 3 $ case.}
  \label{fig:beta10b}
\end{figure}

In order to to compare  SNC with traditional quantum cloning
machines we have to consider a specific system where the required
gates are performed. Obviously this can be done in several
different ways: we choose the $XY$  Hamiltonian as the model system
for both schemes. In particular we compare the two methods for
$M=2$ and $M=3$ equatorial qubits. For the quantum circuit
approach quantum gates are implemented by a time dependent
Hamiltonian. It has been shown~\cite{kempe02,jens} that the $XY$
Hamiltonian is sufficient to implement both one and two-qubit
gates. The elementary two-qubit gate is the iSWAP:
\begin{equation}\label{iswap}
  U(iSWAP)= \left(
\begin{array}[c]{cccc}
1 &  &  & \\
 & 0 & i & \\
 & i& 0 & \\
 &  &  & 1
\end{array}
\right)
\end{equation}
It can be obtained turning on an $XY$ interaction between the two qubits without
external magnetic field and letting them interact for $J t=\pi/4$. By applying the
iSWAP gate twice, the CNOT operation can be constructed
\begin{widetext}
\begin{figure}[htbp]
  \centering
  \[\Qcircuit @C=14pt @R=10pt {
\arr&\ctrl{2}&&  &  \arr& \wire                      &\gate{R_z\left(-\pim\right)}& \multigate{2}{iSWAP}& \gate{R_x\left(\pim\right)}&\multigate{2}{iSWAP}&  \wire&  \\
    &        && =&      &                            &
                                                                                 &      \textrm{iSWAP}& & \textrm{iSWAP}&\\
\arr&\targ   &&  & \arr & \gate{R_x\left(\pim\right)}& \gate{R_z\left(\pim\right)}& \ghost{iSWAP}&\wire&\ghost{iSWAP}&\gate{R_z\left(\pim\right)}&
}
\]
  \caption{Circuit implementing the CNOT from the iSWAP. This circuit
    is used to implement the quantum cloning by means of gates.}
  \label{fig:circuitCNOT}
\end{figure}
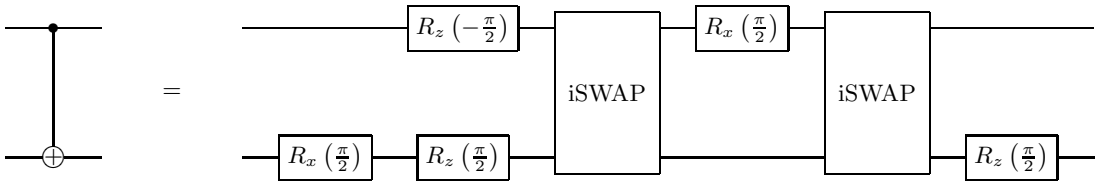
\end{widetext}
This means that we need two two-qubit operations for each CNOT. We
simulated the circuits shown in Fig. \ref{fig:circuitPCC} for
$M=2$ and $M=3$ in the presence of noise and we calculated the
corresponding fidelities. We neglected the effect of noise during
single qubit operations. This is equivalent to assume that the
time needed to perform this gates is much smaller than the typical
decoherence time. The results are shown in Fig.\ref{fig:beta10a}
and Fig.\ref{fig:beta10b}. The fidelity for the quantum gates
(squares) and that for the SNC (circles) are compared as functions of
the coupling parameter $\alpha$.
Even for small $\alpha$ the fidelity for the circuits is much
worse than that for the network. Notice that for $M=3$, though without noise ($\alpha=0$) the SNC fidelity is lower than the ideal one, for $\alpha > \alpha^*=2.5
\cdot 10^{-3}$ the situation is reversed. This shows that our scheme is more
efficient than the one based on quantum gates. Moreover for $M>3$
the time required for quantum circuit PCC grows with increasing M
while, as discussed previously, the optimal $t^{(M)}$ of the SNC
decreases with $M$. This suggests that our proposal is even more
efficient for growing $M$. Changing the model does not affect these results. Indeed the time required
to perform a CNOT using Heisenberg or Ising interactions is just
half the time required for the $XY$ model.

We also believe that in a real implementation the effect of noise on our system
can be very small compared to the that acting on a quantum circuit. This is because
during the evolution the spin network can be isolated from the environment.

\section{The universal cloner with spin networks}
\label{sec:univ} It would be desirable to implement also a
universal quantum cloner by the same method illustrated here. In
this section we briefly report our attempt to implement the $1\to
2$ universal cloner. In the previous sections we demonstrated that
for the models presented the fidelity is invariant on $\varphi$
(phase covariance) but still depends on $\vartheta$. This axial
symmetry relies on the selection of the $z$-axis for the
initialization of the blank spins. In order to perform a universal
cloner we need a spherical symmetry. This means that both the
Hamiltonian and the initial state must be isotropic. The first
condition is fulfilled using the Heisenberg interaction without
static magnetic field that would break the spherical symmetry. The
second requirement can be obtained using for the initial state of
the blank qubits a completely random state. In other words the
complete state of the network (initial state + blanks) is
$$
\rho(0) = \ket{\psi}\bra{\psi}\otimes \frac{1}{4}\one \;\; .
$$
The maximum fidelity is obtained for $J t=\frac{2\pi}{3}$ and has the value
$$
\fid=13/18\simeq 0.72
$$
that has to be compared with the value $5/6 \simeq 0.83$ of the optimal
universal cloner~\cite{BH96}.
Our model is the most general time independent network containing three spins and fulfills
the required conditions.

\section{Quantum cloning of qutrits and qudits}
\label{sec:qudits} 
\begin{figure}[htbp]
\begin{center}
\includegraphics[scale=0.5]{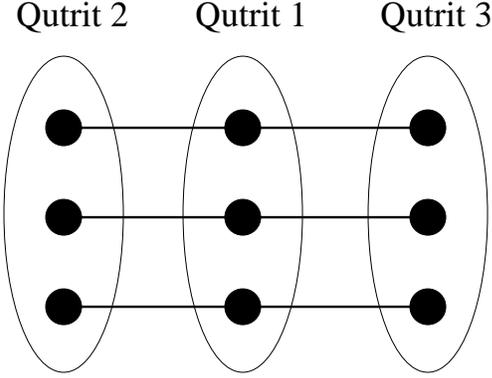}
\end{center}
  \caption{Arrangement of the network for qutrits. Each dot represents 
a spin and an ellipse encloses each logical qutrit. A line connecting 
two dots means that the corresponding spins interact via $XY$ model.}
  \label{fig:qutritxy}
\end{figure}

Spin network cloning technique can be generalized to qutrits and qudits.  
This is what we discuss in this Section starting, for
simplicity, with the qutrit case. The cloning of qudits is a
straightforward generalization. Our task is to find an interaction
Hamiltonian between qutrits able to generate a time evolution as
close as possible to the cloning transformation. One obvious
generalization of the qubit case is to consider qutrits as spin-1
systems. In this picture the three basis states could be the
eigenstates of the angular momentum with z component (-1,0,1). The
natural interaction Hamiltonian would then be the Heisenberg or
the $XY$ interaction
\begin{equation}
  H_I = J_{ij}\vec{S^i}\cdot\vec{S^j} \quad \textrm{or} 
  \quad H_I = J_{ij}(S^i_XS^j_X+S^i_YS^j_Y)
\end{equation}
Alternatively one can think to use the state of physical qubits to
encode the qutrits. Such an encoding, originally proposed in a
different context~\cite{kempe01}, uses three qubits to encode one
single logical qutrit:
\begin{eqnarray*}
  \ket{0}_L&=& \ket{001} \\
    \ket{1}_L &=& \ket{010} \\
   \ket{2}_L &=& \ket{100}
\end{eqnarray*}
In Ref.\cite{kempe02} it is shown that this encoding, together
with a time-dependent $XY$ interaction, is universal for quantum
computing with qutrits. In our work however we have restricted
ourselves to the use of time-independent interactions with a
suitable design of the spin network. For the qubit case the $XY$
interaction is able to swap two spins. We know that this is the
key to clone qubits and so one could try a similar approach also
for qutrits. However, for higher spin, Hund's rule forbids the
swapping. For this reason we have turned our attention to the
encoded qutrits to see if swapping is possible. It is simple to
show that the network depicted in Fig.\ref{fig:qutritxy} satisfies our 
requirements.
In the arrangement each dot represents a spin and three dots
inside an ellipse correspond to an encoded qutrit. A static magnetic 
field $\Delta$ pointing in the $z$ direction is applied to the first spin.  
A line connecting two dots means that they interact via an $XY$ interaction
with amplitude J. It can be easily checked that for a single
couple of qutrits the exchange processes are possible. This
network is the generalization of the spin star that we analyzed
before in which a single qutrit interacts with the others. It is
easily generalized for the $1\to M$ case using three spin stars.
The single qutrit Hamiltonian is realized applying magnetic fields
to the physical qubits.

In analogy with the qubit cloner we will prepare qutrit 1 in the original 
state 
\begin{equation}
  \label{qutritinitial}
  \ket{\psi} = \alpha \ket{0}_L+\beta\ket{1}_L+\gamma\ket{2}_L
\end{equation}
and initialize the other qutrits in a blank state, for example
$\ket{0}_L$.
 Now due to the interactions the state
will evolve in a restricted subspace of the Hilbert space:
\begin{eqnarray}
  \label{qutritpsit}
  \ket{\psi(t)} &=& \alpha \ket{000}_L
+\beta_1\ket{100}_L+\beta_2\ket{010}_L+\beta_3\ket{001}_L \nonumber \\
&+&\gamma_1\ket{200}_L+\gamma_2\ket{020}_L+\gamma_3\ket{002}_L
\end{eqnarray}
To find the
fidelity of the clones with respect to the state of Eq.\eqref{qutritinitial}
we need the reduced density matrix of one of the clones (for example
the third). The result, in the basis $(\ket{0}_L,\ket{1}_L,\ket{2}_L)$,
is
\begin{equation}
  \label{qutredmat}
  \rho_3=
  \begin{pmatrix}
    1-\abs{\beta_3}^2-\abs{\gamma_3}^2& \alpha \beta_3^*&
    \alpha\gamma_3^*\\
    \alpha^*\beta_3& \abs{\beta_3}^2& \beta_3\gamma_3^*\\
    \alpha^*\gamma_3& \beta_3^*\gamma_3&\abs{\gamma_3}^2
  \end{pmatrix}
\end{equation}
In order to find the coefficients $\beta_i(t)$ and $\gamma_i(t)$ we have to
diagonalize the Hamiltonian.
We consider the double PCC of Eq. \eqref{doublePCC}:
 our model is automatically invariant on $\varphi_i$  because there is 
no preferred direction in the space of the
qutrits.
The maximum fidelity achievable with SNC is:
\begin{equation}
  \mathcal F_{3}=\frac{4+2\sqrt{2}}{9}\simeq 0.759
\end{equation}
This value has been obtained with $\Delta/J =1/\sqrt{2}$ and $J
t=\pi/\sqrt 2$. Note that this value is very close to the optimal one and
the difference is only $2\cdot 10^{-3}$.

\begin{figure}[htbp]
  \centering
 \includegraphics*[scale=0.3]{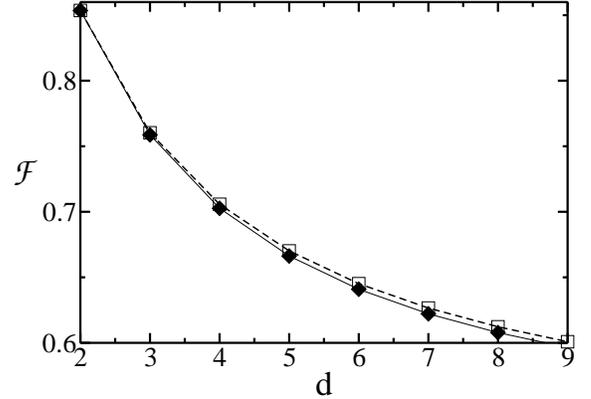}
        \caption{The optimal (square) and the SNC (diamond) fidelities for
    $1\to 2$ PCC in $d$ dimensions are compared.}
      \label{fig:qudit}
\end{figure}

We calculated also the fidelity for the $1\to M$ cloning of qutrits using the
star configuration. The maximum fidelity is:
\begin{equation}
\fid=\frac{2+4\sqrt M+3 M}{9M}
\end{equation}
obtained for the same value of the star configuration of qubits
($J t^{(M)}= \pi/\sqrt M$ and $B^{(M)}/J=\sqrt M /2$).

The generalization to qudits is straightforward. Following the
same approach we encode qudits using $d$ qubits to encode each
qudit. After some algebra one finds the general expression for the
PCC in d dimensions.
The values $t^{(M)}$ and $B^{(M)}$ are independent from $d$ and the 
expression for the fidelity is:
\begin{equation}
  \mathcal F_{1\to  2,d}= \frac{(d-1)(d+2\sqrt{2})+2}{2d^2}
\end{equation}
In Fig.\ref{fig:qudit} the optimal and SNC fidelities are compared.
As we can see, the fidelity of the 
	spin network
    implementation is very close to the ideal one.

\section{Implementation with Josephson nanocircuits}
\label{sec:implementation} 
The final section of this work is
devoted to the possibility of implementing spin network cloning in
solid-state devices. Besides the great interest in solid state
quantum information, nanofabricated devices offer great
flexibility in the design and allow to realize the graphs
represented in Fig.\ref{fig:network}. We analyze the implementation with 
Josephson nanocircuits which are currently considered among the most
promising candidates as building blocks of quantum information
processors~\cite{schoenreview,averinreview}. Here we discuss only the $1
\to 2$ cloning for qubits. The generalization to the other cases is
straightforward.

\begin{figure}[htbp]
  \centering
\includegraphics*[scale=0.6]{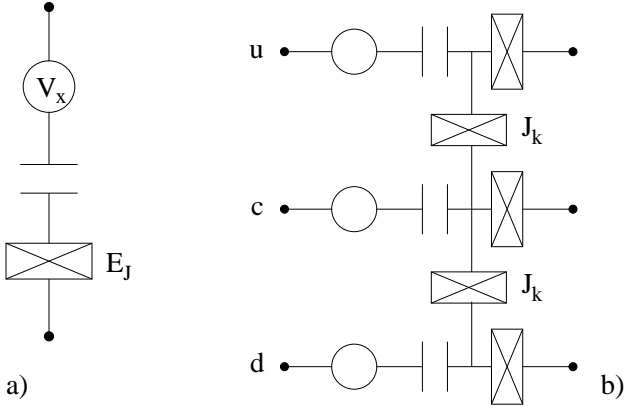}
\caption{a) A sketch of the charge qubit. It consists of a 
	superconducting electron box formed with an applied gate voltage $V_x$. 
	The device operates in the charging regime, i.e. the Josephson couplings $E_J$ 
	of the junction (crossed box in the figure) is much smaller 
	than the charging energy.  
	b) Implementation of the $1 \to 2$ spin network cloning by means of Josephson 
 	qubits. The unknown state to be cloned in stored in the central qubit $c$  while 
        the blank qubits $u$ and $d$ are the ones where the state is cloned. The 
	coupling between the qubits is via the Josephson junctions of coupling energy $J_K$.}
      \label{fig:squbit}
\end{figure}

In the charge regime a Josephson qubit can be realized using a Cooper pair 
box~\cite{schoenreview} (see Fig.\ref{fig:squbit}a), the logical state is 
characterized by the box having zero or one excess charge. Among the various ways to 
couple charge qubits, in order to implement SNC the qubits should be coupled
via Josephson junctions~\cite{ourJLTP} (see Fig.\ref{fig:squbit}b).
The central qubit (denoted by $c$ in the figure) will encode the state to be 
cloned while the upper and lower qubits (denoted with $u$=up and $d$=down) are 
initially in the blank state. All the Josephson junctions are assumed to be tunable 
by local magnetic fluxes. The total Hamiltonian of the 3-qubit system is given by 
the sum of the Hamiltonians of the qubits $H_0$ plus the interaction between them $H_{cou}$.
\begin{equation}
	H_0= \sum_{i=u,c,d} \delta E_c \sigma_z^{(i)} -E_J \sigma_x^{(i)}
\end{equation} 
where $E_J$ is the Josephson coupling in the Cooper pair box  and 
$\delta E_c$ is the energy difference between the two charge states of the 
computational Hilbert space. 
The coupling Hamiltonian for the $3$-qubit system is
\begin{eqnarray}
    H_{cou}  &=&  \sum_{i=u,d} E_K^{(i)} \sigma_z^{(c)}\sigma_z^{(i)}
    \nonumber \\
       &-& (1/2)  \sum_{i=u,d} J_K^{(i)}
                 [\ \sigma^{(c)}_+ \sigma^{(i)}_- \ +\ \ {\rm h.c.\/}\ ]
\label{3clone}
\end{eqnarray}
Here $J_K$ is the Josephson energy of the junctions which couple the different 
qubits and $\sigma_\pm=(\sigma_x\pm i\sigma_y)/2$.
If the coupling capacitance between the qubits is very small as compared to the
other capacitances one can assume $E_K^{(j)}$ to be negligible.
In practice, however the capacitive coupling is always present therefore it is 
necessary to have $J_K^{(j)}\gg 4E_K^{(j)}$.
Then the dynamics of the system approximates the ideal $XY$ dynamics required
to perform quantum cloning. The protocol to realize the SNC requires the preparation 
of the initial state. This can be achieved by tuning the gate voltages in such a 
way that the blank qubits are in $\ket 0$ and the central qubit is in the state to 
be cloned. During the preparation the coupling between the qubits should be kept 
zero by piercing the corresponing SQUID loops of the junctsion $J_k$ with a magnetic 
field equal to a flux quantum. In the second step, $H_0$ is switched off and the 
dynamics of the system is entirely governed by $H_{cou}$. At the optimal time the 
original state is cloned in the $u$ and $d$ qubits.

\begin{figure}[htbp]
  \centering
\includegraphics*[scale=0.3]{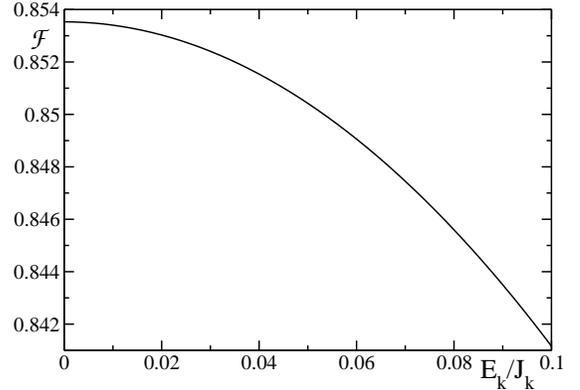}
\caption{ Fidelity for the coupling Hamiltonian \eqref{3clone} 
	as a function of $E_K/J_K$.}
      \label{fig:fidelityreal}
\end{figure}

As the implementation with superconducting nanocircuits has a slightly different 
Hamiltonian as compared to the ideal $XY$ model it is important to check for the 
loss of fidelity due to this difference.
As it is shown in Fig.\ref{fig:fidelityreal}, for $J_K/E_K \le 0.1$ the maximum 
fidelity achievable differs at most by $ \sim 10^{-2}$ from the ideal value.

\section{Conclusions}
\label{sec:conclusion}
We have demonstrated that quantum cloning, in particular PCC, can
be realized using no external control but just with an appropriate
design of the  system Hamiltonian. We considered the Heisenberg
and $XY$ coupling between the qubits and we found that the $XY$ model
saturates the optimal value for the fidelity of the $1\to 2$ PCC.
In all other cases we have analyzed ($N \to M $ PCC, universal
cloning, cloning of qudits) our protocol gives a value of the
fidelity of clones that is always within a few percent
 of the optimal value. As compared to the standard protocol using quantum
gates, however, there is a major advantage. Our setup is fast and,
moreover, its execution time does not increase with the number of
qubits to be cloned. In the presence of noise this allows to reach
a much better fidelity than the standard protocol even in the
presence of a weak coupling to the external environment. In
addition we expect that the system in the SNC is better isolated
from the external environment because no gate pulses are needed.
Finally we proposed a possible implementation of our scheme using
superconducting devices available with present day technology.
This would be the first experimental realization of quantum
cloning in solid state systems. We want to stress that our results
on cloning together with others on communication and computation
open new perspectives in the realization of a quantum processor,
reducing the effect of noise on the system. It would be
interesting to consider if it is possible to realize other quantum
information protocols or quantum algorithms, using time
independent spin networks.

This work was supported by the European Community under contracts IST-SQUIBIT,
IST-SQUBIT2, IST-QUPRODIS, IST-SECOQC, and RTN-Nanoscale Dynamics.



\begin{thebibliography}{99}

\bibitem{WZ82}
    W.K. Wootters and W.H. Zurek, Nature {\bf 299}, 802 (1982).
\bibitem{gisin02}
    N. Gisin and G. Ribordy and W. Tittel and H. Zbinden,
    Rev. Mod. Phys. {\bf 74}, 145 (2002).
\bibitem{BH96}
    V. Bu\v{z}ek and M. Hillery, Phys. Rev. A {\bf 54}, 1844 (1996).

\bibitem{galvao}
    E.F. Galv\~{a}o and L. Hardy, Phys. Rev. A {\bf 62}, 022301 (2000).
\bibitem{lopresti}
    G. M. D'Ariano and P. Lo Presti,
    Phys. Rev. A {\bf 64}, 042308 (2001).
\bibitem{bruss98}
        D. Bru{\ss}, D. P. DiVincenzo, A. Ekert, C. A. Fuchs,
        C. Macchiavello, J. A. Smolin, Phys. Rev. A {\bf57}, 2368 (1998).
\bibitem{cloner_n-m}
        N. Gisin and S. Massar, Phys. Rev. Lett. {\bf 79}, 2153-2156 (1997);
        D. Bruss, A. Ekert and C. Macchiavello, Phys. Rev. Lett. {\bf 81},
        2598 (1998); R.~F.~Werner, Phys.~Rev. A{\bf 58}, 1827 (1998).
\bibitem{bruss00}
        D. Bru{\ss}, M. Cinchetti, G. M. D'Ariano, C. Macchiavello,
        Phys. Rev. A {\bf 62}, 012302 (2000).
\bibitem{cummins02}
        H. K. Cummins, C. Jones, A. Furze, N. F. Soffe, M. Mosca,
        J. M. Peach, J. A. Jones, Phys. Rev. Lett. {\bf 88}, 187901
        (2002).
\bibitem{bouwmeester02}
        A. Lama-Linares, C. Simon, J.-C. Howell and D. Bouwmeester,
        Science {\bf 296}, 712 (2002).
\bibitem{demartini}
        D. Pelliccia, V. Schettini, F. Sciarrino, C. Sias and
        F. De~Martini,
        Phys. Rev. A {\bf 68}, 042306 (2003);
        F. De~Martini, D. Pelliccia and F. Sciarrino,
Phys. Rev. Lett. {\bf 92}, 067901 (2004).
\bibitem{ekert03}
        J. Du, T. Durt, P. Zou, L.C. Kwek, C.H. Lai, C.H. Oh, and A.~Ekert,
        quant-ph/0311010.
\bibitem{niu99}
        C.-S. Niu and R.B. Griffiths,  Phys. Rev. A {\bf 60}, 2764 (1999).
\bibitem{benjamin03}
    S. C. Benjamin and S. Bose, Phys. Rev. Lett. {\bf 90}, 0247901 (2003).
\bibitem{yung}
    M.-H. Yung, D.W. Leung and S. Bose, Quantum Inf. Comput. {\bf 4}, 174
(2004).
\bibitem{bose}
    S. Bose, Phys. Rev. Lett. {\bf 91}, 207901 (2003).
\bibitem{subra03}
    V. Subrahmanyam, Phys. Rev. A {\bf 69}, 034304 (2004).
\bibitem{osborne}
    T. J. Osborne and N. Linden, Phys. Rev. A {\bf 69}, 052315 (2004).
\bibitem{datta}
    M. Christandl, N. Datta, A. Ekert,
        and A. J. Landahl,
    Phys. Rev. Lett. {\bf 92}, 187902 (2004).
\bibitem{lloyd}
    S. Lloyd, Phys. Rev. Lett. {\bf 90}, 167902 (2003).
\bibitem{cirac}
    F. Verstraete, M. A. Mart\'in-Delgado,
        and J. I. Cirac,
    Phys. Rev. Lett. {\bf 92}, 087201 (2004).
\bibitem{giovannetti}
    V. Giovannetti and R. Fazio, Phys. Rev. A {\bf 71}, 032314 (2005).
\bibitem{romito}
    A. Romito, R. Fazio and C. Bruder, Phys. Rev. B {\bf 71} 100501 (2005).
\bibitem{paternostro}
    M. Paternostro, M.S. Kim, G.M. Palma, and G. Falci, Phys. Rev. A {\bf 71}, 042311 (2005).
\bibitem{dechiara}
    G. De Chiara, R. Fazio, C. Macchiavello,
    S. Montangero, and G. M. Palma, Phys. Rev. A {\bf 70}, 062308 (2004).
\bibitem{dariano03}
    G. M. D'Ariano and C. Macchiavello, Phys. Rev. A {\bf 67}, 042306 (2003).
\bibitem{olayacastro}
    A. Olaya-Castro, N. F. Johnson, and L. Quiroga, Phys. Rev. Lett. {\bf 94}, 110502 (2005).
\bibitem{footnote}
    There is  not a unique formula for arbitrary $N$.
\bibitem{werner98}
    R. F. Werner, Phys. Rev. A {\bf 58}, 1827 (1998).

\bibitem{gisinqutrit}
    N.J. Cerf, T. Durt and N. Gisin, J. Mod. Opt., {\bf 49}, 1355 (2002).
\bibitem{matsu}
    H. Fan, H. Imai, K. Matsumoto and X.-B. Wang,
    Phys. Rev. A {\bf 67}, 022317 (2003)
\bibitem{bdm} F. Buscemi, G. M. D'Ariano and C. Macchiavello,
Phys. Rev. A {\bf 71}, 042327 (2005).
\bibitem{du}
    J. Du, T. Durt, P. Zou, L.C. Kwek, C.H. Lai, C.H. Oh and
    A. Ekert, Phys. Rev. Lett. {\bf 94}, 040505.
\bibitem{buzek97}
    V. Bu\v{z}ek, S.L. Braunstein, M. Hillery and D. Bru{\ss},
    Phys. Rev. A {\bf 56}, 3446 (1997).
\bibitem{Bosestar}
    A. Hutton and S. Bose, Phys. Rev. A {\bf 69}, 042312 (2002).
\bibitem{fiurasek03}
    J. Fiur\'{a}\v{s}ek, Phys. Rev. A {\bf 67}, 052314 (2003).
\bibitem{cohen}
    C. Cohen-Tannoudji, J. Dupont-Rac and G. Grynberg,
    {\em Atom-Photon Interactions},
    John~Wiley \& Sons, New York, (1992)
\bibitem{kempe02}
    J. Kempe and K.B. Whaley, Phys. Rev. A, {\bf 65}, 052330 (2002).
\bibitem{jens}
    N. Schuch and J. Siewert, Phys. Rev. A  {\bf 67}, 032301 (2003).
\bibitem{kempe01}
    J. Kempe, D. Bacon, D. P. DiVincenzo and K.B. Whaley, in
    {\em Quantum Information and Computation}, R. Clark {\em et al.} Eds.,
    Rinton Press, New Jersey, Vol.1, 33 (2001).
\bibitem{schoenreview}
    Yu. Makhlin, G. Sch\"on, and A. Shnirman,
    Rev. Mod. Phys. {\bf 73}, 357 (2001).
\bibitem{averinreview}
    D.~V. Averin, Fortschr. Phys. {\bf 48}, 1055 (2000).
\bibitem{ourJLTP}
    J.\ Siewert, R.\ Fazio, G.M.\ Palma, and E.\ Sciacca,
        J.\ Low Temp.\ Phys.\ {\bf 118}, 795 (2000).




\end{thebibliography}
\end{document}